\documentclass[12pt,a4paper,oneside,final,tilepage,onecolumn,thmsal]{article}
\usepackage[dvips]{graphics}
\usepackage{epsfig}
\usepackage{amsmath}

\newcommand{\QED}{\hspace*{\fill}\rule{2.5mm}{2.5mm}}

\usepackage{graphics}
\begin{document}
\def\beq{\begin{equation}}
\def\eeq{\end{equation}}
\def\bea{\begin{eqnarray}}
\def\eea{\end{eqnarray}}
\def\ve{\vert}
\def\vel{\left|}
\def\ver{\right|}
\def\nnb{\nonumber}
\def\ga{\left(}
\def\dr{\right)}
\def\aga{\left\{}
\def\adr{\right\}}
\def\rar{\rightarrow}
\def\nnb{\nonumber}
\def\la{\langle}
\def\ra{\rangle}
\def\ba{\begin{array}}
\def\ea{\end{array}}
\def\ds{\displaystyle}
\title{{\small {\bf Exact solutions for vibrational levels of the Morse potential via
the asymptotic iteration method }}}
\author{\vspace{1cm}\\
{\small T. Barakat\thanks {electronic address:
zayd95@hotmail.com} , and K. Abodayeh} \\ {\small Physical Sciences Department, Prince Sultan University},\\
{\small Riyadh 11586, Saudi Arabia} }
\date{}
\begin{titlepage}
\maketitle
\thispagestyle{empty}
\begin{abstract}
\baselineskip .8 cm Exact solutions for vibrational levels of
diatomic molecules via the Morse potential are obtained by means
of the asymptotic iteration method. It is shown that, the
numerical results for the energy eigenvalues of $^{7}Li_{2}$ are
all in excellent agreement with the ones obtained before. Without
any loss of generality, other states and molecules could be
treated in a similar way.
\end{abstract}
\vspace{1cm} PACS number(s): 03.65.Ge
\end{titlepage}
\section{{\small Introduction}}
\baselineskip .8cm \hspace{0.6cm}Since the appearance of the
Schr\"{o}dinger equation in quantum mechanics, there have been
continual researches for studying Schr\"{o}dinger equation with an
exactly solvable potentials by using differenet methods. The
range of potentials for which Schr\"{o}dinger equation can be
solved exactly has been extended considerably owing to the
investigations inspired, for example, by super-symmetric quantum
mechanics [1], shape invariance [2], the factorization method
[3-8], and recently the asymptotic iteration method (AIM) [9].

In recent years much attention has been focused on AIM. This method
reproduces exact solutions to many differential equations which are
important in applications to many problems in physics, such as the
equations of Hermite, Laguerre, Legendre, and Bessel [9]. The AIM
also gives a complete exact solutions of Schr\"{o}dinger equation
for P\"{o}sch-Teller potential, the harmonic oscillator potential,
the complex cubic, quartic [10], and sextic anharmonic oscillator
potentials [11, 12]. Very recently we applied the AIM and found the
exact eigenvalues for the angular spheroidal wave equation [13, 14].

Encouraged by its satisfactory performance through comparisons
with the other methods, we feel tempted to extend AIM to solve
exactly the one-dimensional Schr\"{o}dinger equation with the
Morse potential [15]. This potential has played an important role
in many different fields of physics such as molecular physics,
solid state physic, and chemical physics, etc. This potential has
been studied by many different approaches such as the standard
confluent hypergeometric functions [16], the algebraic method
[17], the supersymmetric method [18], the coherent states [19],
the controllability [20], the series solutions with the mass
distribution [21], laplace transforms [22], etc.

Fortunately, the one-dimensional Schr\"{o}dinger equation with the
Morse potential can be reduced to a second-order homogenous
linear differential equation, and we therefore directly can make
use of AIM to formulate an elegant algebraic approach to yield a
fairly simple analytic formula which gives rapidly the exact
energy eigenvalues with high accuracy. Most importantly, the
numerical computation of the Morse potential energy eigenvalues
using this method is quite simple, fast, and the energy
eigenvalues were satisfying a simple ordering relation.
Therefore, one can unambiguously select the correct starting
energy eigenvalue. Moreover, this work shows that, the AIM can be
a simple alternative approach for computing the vibrational
levels of the Morse potential.

 In this spirit, this paper is organized as follows. In Sec. 2 the
 asymptotic iteration method
for Schr\"{o}dinger equation with the Morse potential is
outlined. The analytical expressions for asymptotic iteration
method are cast in such a way that allows the reader to use them
without proceeding into their derivation. In Sec. 3 we present
our numerical results compared with other works, and then we
conclude and remark therein.

\section{{\small Formalism of the asymptotic iteration method for
Schr\"{o}dinger equation with the Morse potential}}
 \hspace{0.6cm}As an empirical potential, the Morse potential has been one of
the most useful and convenient model, which gives an excellent
qualitative description of the interaction between two atoms in a
diatomic molecule. As we know the rotational energy of a molecule
is much smaller than that of its vibrational energy, and
therefore, in a pure Morse potential model the rotational energy
of a molecule has been omitted. Hence, the Schr\"{o}dinger
equation for Morse potential with angular momentum $\ell=0$ is
\begin{eqnarray}
\left[-\frac{\hbar^2}{2\mu}\frac{d^{2}}{dx^{2}}+V(x)\right]\Psi_{n}(x)=E_{n}\Psi_{n}(x),
\end{eqnarray}
where $E_{n}$ are the energy eigenevalues, and $V(x)$ is the
Morse potential function
\begin{eqnarray}
V(x)=D_{e}(e^{-2\beta(x-x_{e})}-2e^{-\beta(x-x_{e})}).
\end{eqnarray}

Equation (1) is called one-dimensional schr\"{o}dinger equation,
if $x$ is defined on the whole line ($-\infty< x <+\infty$), and
the eigenfunctions are normalized
$\int^{+\infty}_{-\infty}|\Psi_{n}(x)|^2dx=1$. However, for real
diatomic molecules $x$ should ranges from $0$ to $\infty$.

$D_{e}$ is the dissociation energy, $x_{e}$ is the equilibrium
internuclear distance of a diatomic molecules, $\mu$ is the reduced
mass, and $\beta$ is an adjustable parameter. Morse potential has a
minimum value at $x=x_{e}$, and it is zero at $x=\infty$. At $x=0$,
$V(0)$ has a finite value of $D_{e}(e^{2\beta x_{e}}-2e^{\beta
x_{e}})$ that is positive when $\beta x_{e}> ln2$.

Starting with Morse's substitution
$u=e^{\frac{-\beta(x-x_{e})}{2}}$, we rewrite equation (1) in the
form
\begin{eqnarray}
-\frac{d^{2}\Psi_{n}(u)}{du^{2}}-\frac{1}{u}\frac{d\Psi_{n}(u)}{du}+\frac{8\mu
D_{e}}{\beta^{2}\hbar^{2}}[u^{2}-2]\Psi_{n}(u)=\frac{8\mu
E_{n}}{\beta^{2}\hbar^{2}u^{2}}\Psi_{n}(u).
\end{eqnarray}
Furthermore, we remove the first derivative by proposing the
ansatz
\begin{eqnarray}
\Psi_{n}(u)=\Phi_{n}(u)exp(-p(r)/2);~~~~p^{'}(r)=\frac{1}{u}.
\end{eqnarray}
Which in turn implies
\begin{eqnarray}
-\frac{d^{2}\Phi_{n}(u)}{du^{2}}+\frac{\epsilon_{n}(\epsilon_{n}+1)}{u^{2}}\Phi_{n}(u)+\gamma^{2}u^{2}\Phi_{n}(u)
=2\gamma^{2}\Phi_{n}(u),
\end{eqnarray}

where
\begin{eqnarray}
\epsilon_{n}(\epsilon_{n}+1)=-\frac{1}{4}-\frac{8\mu
E_{n}}{\beta^{2}\hbar^{2}};~~~~\gamma^{2}=\frac{8\mu
D_{e}}{\beta^{2}\hbar^{2}}.
\end{eqnarray}
If we further introduce the frequency
\begin{eqnarray}
\omega_{0}=\beta\sqrt{2D_{e}/\mu}
\end{eqnarray}
of classical small vibrations about the equilibrium position
$x=x_{e}$, and express the energy parameters in unit
$\hbar\omega_{0}$; that is,
\begin{eqnarray}
D_{e}=\Delta\hbar\omega_{0};~~~~
E_{n}=\varepsilon_{n}\hbar\omega_{0};~~~~\varepsilon_{n}=-\frac{(\epsilon_{n}+\frac{1}{2})^{2}}{16\Delta^{2}},
\end{eqnarray}
it is now more convenient to write the eigenvalue problem in the
re-scaled form
\begin{eqnarray}
-\frac{d^{2}\Phi_{n}(u)}{du^{2}}+\frac{\epsilon_{n}(\epsilon_{n}+1)}{u^{2}}\Phi_{n}(u)+16\Delta^{2}u^{2}\Phi_{n}(u)
=32\Delta^{2}\Phi_{n}(u).
\end{eqnarray}
In order to guarantee the asymptotic behaviour of this eigenvalue
problem when $u\longrightarrow\infty$, and~ $u\longrightarrow 0$
we found that this asymptotic behaviour suggests that
$\Phi_{n}(u)$ should look like
\begin{eqnarray}
\Phi_{n}(u)=u^{(\epsilon_{n}+1)}e^{-2\Delta u^2}f_{n}(u).
\end{eqnarray}

This implies that the function $f_{n}(u)$ will satisfy a
second-order homogenous linear differential equation of the form
\begin{eqnarray}
\frac{d^2f_{n}(u)}{du^2}-(8\Delta u-\frac{2\epsilon_{n}+2}{u})
\frac{df_{n}(u)}{du}-(12\Delta+8\Delta\epsilon_{n}-32\Delta^{2})f_{n}(u)=0.
\end{eqnarray}

The systematic procedure of the asymptotic iteration method
begins now by rewriting equation (11) in the following form
\begin{eqnarray}
f_{n}^{''}(u)=\lambda_{0}(u)f_{n}^{'}(u)+s_{0}(u)f_{n}(u),
\end{eqnarray}
where
\begin{eqnarray}
\lambda_{0}(u)=(8\Delta u-\frac{2\epsilon_{n}+2}{u}),~ {\rm and}
~~~~s_{0}(u)=(12\Delta+8\Delta\epsilon_{n}-32\Delta^{2}).
\end{eqnarray}
The primes of $f_{n}(u)$ in equation (12) denote derivatives with
respect to $u$.

 Now, in order to find a general solution to this equation we rely
on the symmetric structure of the right hand side of equation
(12). Thus, if we differentiate equation (12) with respect to
$u$, we obtain
\begin{eqnarray}
f_{n}^{'''}(u)=\lambda_{1}(u)f_{n}^{'}(u)+s_{1}(u)f_{n}(u),
\end{eqnarray}
where

$\lambda_1(u)=\lambda^{'}_{0}(u)+s_{0}(u)+\lambda^{2}_{0}(u)$,~
 and~~~ $s_{1}(u)=s^{'}_{0}(u)+s_{0}(u)\lambda_{0}(u)$.

Likewise, the calculations of the second derivative of equation
(14) yield
\begin{eqnarray}
f_{n}^{''''}(u)=\lambda_{2}(u)f_{n}^{'}(u)+s_{2}(u)f_{n}(u),
\end{eqnarray}
where

$\lambda_{2}(u)=\lambda^{'}_{1}(u)+s_{1}(u)+\lambda_{0}(u)\lambda_{1}(u)$,
~and~~~
 $s_{2}(u)=s^{'}_{1}(u)+s_{0}(u)\lambda_{1}(u)$.

For $(k + 1)^{th}$, and $(k + 2)^{th}$ derivatives, $k = 1, 2, . .
.$, one can obtain
\begin{eqnarray}
f_{n}^{(k+1)}(u)=\lambda_{k-1}(u)f_{n}^{'}(u)+s_{k-1}(u)f_{n}(u),
\end{eqnarray}
and
\begin{eqnarray}
f_{n}^{(k+2)}(u)=\lambda_{k}(u)f_{n}^{'}(u)+s_{k}(u)f_{n}(u),
\end{eqnarray}
respectively, where
\begin{eqnarray}
\lambda_{k}(u)=\lambda^{'}_{k-1}(u)+s_{k-1}(u)+\lambda_{0}(u)\lambda_{k-1}(u),
~{\rm and}~~~ s_{k}(u)=s^{'}_{k-1}(u)+s_{0}(u)\lambda_{k-1}(u).
\end{eqnarray}
The ratio of the $(k + 2)^{th}$, and $(k + 1)^{th}$ derivatives,
can be expressed as:
\begin{eqnarray}
\frac{d}{du}ln(f_{n}^{(k+1)}(u))=\frac{f_{n}^{(k+2)}
(u)}{f_{n}^{(k+1)} (u)}=
\frac{\lambda_{k}(f_{n}^{'}(u)+\frac{s_{k}(u)}{\lambda_{k}(u)}f_{n}
(u))}{\lambda_{k-1}(f_{n}^{'} (u)+
\frac{s_{k-1}(u)}{\lambda_{k-1}(u)}f_{n} (u))}.
\end{eqnarray}
For sufficiently large $k$, we can now introduce the "asymptotic"
aspect of the method; that is,
\begin{eqnarray}
\frac{s_k(u)}{\lambda_k(u)}=\frac{s_{k-1}(u)}{\lambda_{k-1}(u)}\equiv\varrho(u).
\end{eqnarray}
Thus equation (19) can be reduced to
\begin{eqnarray}
\frac{d}{du}ln(f_{n}^{(k+1)}
(u))=\frac{\lambda_{k}(u)}{\lambda_{k-1}(u)},
\end{eqnarray}
which yields
\begin{eqnarray}
f_{n}^{(k+1)}
(u)=C_{1}exp\left(\int\frac{\lambda_{k}(u)}{\lambda_{k-1}(u)}du\right)=
C_{1}\lambda_{k-1}(u)exp\left(\int(\varrho(u)+\lambda_{0}(u))du\right),
\end{eqnarray}
where $C_{1}$ is the integration constant, and the right hand side
of equation (22) follows from equation (18), and the definition of
$\varrho$. Substituting equation (22) into equation (16) we obtain a
first-order differential equation
\begin{eqnarray}
f_{n}^{'}(u)+\varrho(u) f_{n}(u)
=C_{1}exp\left(\int(\varrho(u)+\lambda_{0}(u))du\right),
\end{eqnarray}
which, in turn, yields the general solution to equation (12)
\begin{eqnarray}
f_{n}(u)=exp\left(-\int^{u}\varrho(u^{'})
du^{'}\right)\left[C_{2}+C_{1}\int^{u}
exp\left(\int^{u^{'}}\{\lambda_{0}(u^{''})+2\varrho(u^{''})\}du^{''}\right)du^{'}\right].
\end{eqnarray}
Here, it should be noted that one can construct the
eigenfunctions $f_{n}(u)$ from the knowledge of $\varrho$.
\section{{\small Numerical results for the vibrational levels of the Morse potential }}
Within the framework of the AIM mentioned in the above section,
the energy eigenvalues of the Morse potential $\varepsilon_{n}$
are calculated by means of equation (20). To obtain the energy
eigenvalues $\varepsilon_{n}$, first equation (20) is solved for
$\epsilon_{n}$ where the iterations should be terminated by
imposing a condition $\delta_{n}(u)$= 0 as an approximation to
equation (20). On the other hand, for each iteration, the
expression
$\delta_{n}(u)=s_{n}(u)\lambda_{n-1}(u)-s_{n-1}(u)\lambda_{n}(u)$
 depends on two variables: $\epsilon_{n}$, and $u$. The
calculated $\epsilon_{n}$ by means of this condition should,
however, be independent of the choice of $u$. Nevertheless, the
choice of $u$ is observed to be critical only to speed of the
convergence to $\epsilon_{n}$, as well as for the stability of
the process. In this work it is observed that, the best starting
value for $u$ is the value at which the effective potential of
equation (9) takes its minimum value, that is when $u=1$.
Therefore, at the end of the iterations we put $u=1$.

The results of the AIM for $\epsilon_{n}$ with different values
of $n$, yield

\begin{eqnarray}
\epsilon_{0}=\frac{-3+8\Delta}{2},~\epsilon_{1}=\frac{-7+8\Delta}{2},~
\epsilon_{2}=\frac{-11+8\Delta}{2}, .........
\end{eqnarray}
respectively, that means
\begin{eqnarray}
\epsilon_{n}=\frac{-4n-3+8\Delta}{2},~{\rm for}~~
n=0,1,2,.........
\end{eqnarray}
The parameters $\epsilon_{n}$ were calculated by means of 18
iterations only. Therefore, the exact energy eigenvalues of the
Morse potential $\varepsilon_{n}$ are
\begin{eqnarray}
\varepsilon_{n}=-\frac{(-2(2n+1)+8\Delta)^2}{64\Delta},~{\rm
for}~~ n=0,1,2,.........
\end{eqnarray}

For numerical illustration, in table I we calculate the
vibrational energies of the $^{7}Li_{2}$ molecule in the
$A^{1}\Sigma^{+}_{u}$ electronic state. The parameters of the
respective Morse potential are explicitly indicated, with the
dissociation energy parameter in both units $cm^{-1}$, and
$\hbar\omega_{0}$; the later is better suited to compare with the
unit separation of the corresponding levels of the harmonic
oscillator and appreciate the decreasing separations in the Morse
potential. The table includes the energy eigenvalues according to
Morse's exact solution, E. Ley-Koo et al. [23], and to the
calculations of this work using AIM. One can also compare the
results of this work with those of  H. Ta\c{s}eli [24], and can
easily judge the accuracy of the AIM.

Finally, we would like to emphasize that, within the framework of
the AIM, we have easily obtained the exact bound state solutions
for the one-dimensional Schr\"{o}dinger equation for the Morse
potential.

\clearpage
\begin{table}
\begin{center}
\caption{Energy eigenvalues of Morse potential for $^{7}Li_{2}$
in the $A^{1}\Sigma_{u}^{+}$ state with $D_{e}= 8940~ cm^{-1} =
34.997 ~\hbar \omega_{0}$, $x_{0}=3.10821$, and $\beta=0.616$.}
\vspace{1cm}
\begin{tabular}{cccccc}
\hline \hline
$n$&$\varepsilon_{n}$(Morse) [15]& $\varepsilon_{n}$ [23]& $\varepsilon_{n}$(AIM)\\
\hline
0   &-34.4987869262695313     &-34.4987858673604677     &-34.4987858673600556    \\
1   &-33.5130744534329139     &-33.5130728062414320     &-33.5130728062405367     \\
2   &-32.5416483298483712     &-32.5416466840021528     &-32.5416466840014849     \\
3   &-31.5845091444715536     &-31.5845075006434506     &-31.5845075006429106     \\
4   &-30.6416568973024646     &-30.6416552561655102     &-30.6416552561648174      \\
5   &-29.7130915883411006     &-29.7130899505676886     &-29.7130899505671913      \\
6   &-28.7988132175874689     &-28.7988115838512790     &-28.7988115838500462      \\
7   &-27.8988217850415658     &-27.8988201560141817     &-27.8988201560133717      \\
8   &-27.0131172907033879     &-27.0131156670575088     &-27.0131156670571784       \\
9   &-26.1416997345729385     &-26.1416981169820595     &-26.1416981169814555       \\
10  &-25.2845691166502178     &-25.2845675057870984     &-25.2845675057862103        \\
11  &-24.4417254369352221     &-24.4417238334716096     &-24.4417238334714426         \\
12  &-23.6131686954279552     &-23.6131671000377494     &-23.6131671000371455        \\
13  &-22.7988988921284204     &-22.7988973054838091     &-22.7988973054833259        \\
14  &-21.9989160270366071     &-21.9989144498102718     &-21.9989144498099840        \\
15  &-21.2132201001525260     &-21.2132185330174288     &-21.2132185330171161        \\
16  &-20.4418111114761700     &-20.4418095551052090     &-20.4418095551047223       \\
17  &-19.6846890610075462     &-19.6846875160729802     &-19.6846875160728061       \\
18  &-18.9418539487466475     &-18.9418524159217014     &-18.9418524159213639      \\
19  &-18.2133057746934739     &-18.2133042546505699     &-18.2133042546503994      \\
20  &-17.4990445388480325     &-17.4990430322602215     &-17.4990430322599053      \\
21  &-16.7990702412103161     &-16.7990687487502584     &-16.7990687487498924       \\
22  &-16.1133828817803284     &-16.1133814041206413     &-16.1133814041203536       \\
23  &-15.4419824605580693     &-15.4419809983712994     &-15.4419809983712888       \\
24  &-14.7848689775435389     &-14.7848675315028206     &-14.7848675315027016       \\

 \hline\hline
\end{tabular}
\end{center}
\end{table}
\end{document}